# Integrating solid direct air capture systems with green hydrogen production: Economic synergies of sector coupling

Sunwoo Kim[1,2], Joungho Park[3], and Jay H. Lee[1*]

[1] Mork Family Department of Chemical Engineering and Material Sciences, University of Southern California, 925 Bloom Walk, Los Angeles, CA 90089, USA

[2] Department of Chemical and Biomolecular Engineering, Korea Advanced Institute of Science and Technology (KAIST), 291, Daehak-ro, Yuseong-gu, Daejeon, 34141, Republic of Korea

[3] Energy AI & Computational Science Laboratory, Korea Institute of Energy Research, 71-2 Jang-dong, Yuseong-gu, Daejeon, 305-343, Republic of Korea

[*] Corresponding author's E-mail: jlee4140@usc.edu

**Abstract**

With escalating atmospheric $CO_2$ levels, the imperative of direct air capture (DAC) systems becomes increasingly clear. Concurrently, green hydrogen (GH) emerges as a pivotal medium for renewable energy. Nevertheless, the substantial costs associated with these technologies impede their widespread adoption, primarily due to significant installation costs and underutilized capacities when deployed independently. Integrating these technologies through sector coupling can enhance system efficiency and sustainability, while shared power sources and energy storage devices offer additional economic advantages. This study explores the opportunity of sector coupling and assesses the economic viability of two prominent GH technologies—proton exchange membrane (PEM) electrolyzers and alkaline electrolyzers—within this integrated strategy. Our findings indicate that combining GH production with solid DAC systems offers significant economic improvements, enhancing cost-effectiveness by approximately 10% for both types of electrolyzers. These economic benefits persist even amid substantial variations in facility costs. Additionally, increasing the operating flexibility of heat pumps by 10% reduces total expenses by about 4%, further enhancing the economic appeal of sector coupling. These results underscore the potential to improve both the efficiency and economic viability of DAC and GH technologies, promoting broader adoption of cleaner energy solutions.

*Keywords:* Solid direct air capture, Green hydrogen, Alkaline and PEM electrolyzer, Sector coupling

# Graphical abstract

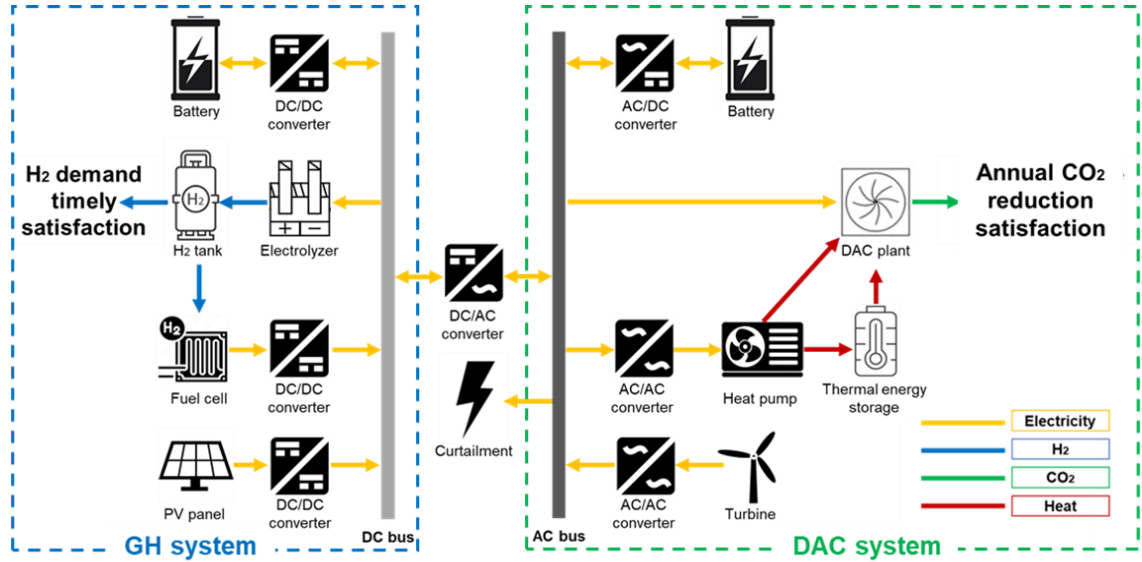

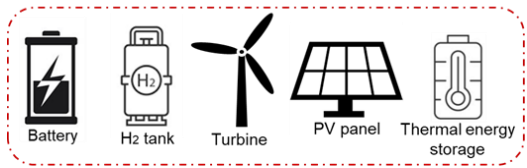

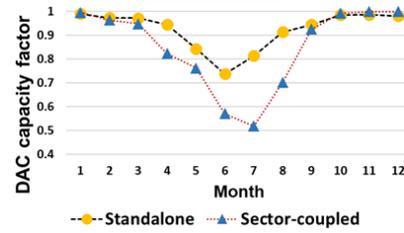

- Reduced total capacity requirement through the sharing of facilities
- Increased variations in the capacity factor of DAC systems

# 1. Introduction

The imperative for a sustainable global society increasingly focuses on mitigating anthropogenic carbon dioxide ($CO_2$) emissions [1-3]. Direct air capture (DAC) systems are gaining recognition as a vital technology for recovering atmospheric $CO_2$ [4-6]. This perspective is supported by a growing consensus among research entities and governments that a transition to clean, renewable energy must be complemented by technologies capable of actively extracting $CO_2$ from the atmosphere [7, 8].

Among the various DAC technologies, low-temperature solid absorbent DAC is emerging as a particularly promising method [9, 10]. Solid DAC technology operates at ~100 °C and utilizes an amine-functionalized sorbent material for $CO_2$ separation via temperature-vacuum swing adsorption/desorption cycles [11]. Compared to high-temperature liquid solvent DAC technology, which employs KOH absorption and regeneration via calcium looping at ~700 °C, solid DAC systems offer superior energy efficiency [12]. During the adsorption phase, $CO_2$ is captured from the atmosphere using solid adsorbents [13]. The captured $CO_2$ is then released for storage during the desorption phase [14]. One notable advantage of solid DAC system is their ability to enter a shutdown state, providing unparalleled adaptability to fluctuating renewable energy availability driven by their modularized nature [15]. However, despite the inherent operational flexibility of solid DAC, the high installation costs of these facilities require them to operate at high loads to achieve economic viability. This often leads to underutilization of their operational flexibility in standalone solid DAC systems.

In tandem with DAC, green hydrogen (GH) is becoming increasingly significant as a sustainable energy carrier [16-20]. GH refers to hydrogen produced by utilizing renewable energy sources, typically through methods like water electrolysis, employing alkaline and proton exchange membrane (PEM) and electrolyzers [21-23]. Alkaline electrolyzers use a liquid electrolyte, typically a 25-30% aqueous KOH-solution, necessitating a higher minimum load by ~20% to maintain stable electrochemical reactions [24]. In contrast, PEM electrolyzers use a solid polymer electrolyte and require noble metal catalyst such as iridium for the anode and platinum for the cathode, which allows efficient proton conduction and a rapid response to varying electrical inputs, have a lower minimum operating threshold by ~5% [24, 25]. Consequently, the adoption of alkaline electrolyzers often hindered due to the considerable expenses associated with substantial storage devices needed to prevent shutdowns, leading to critical degradation. In the net zero emission scenario, the global $H_2$ demand is projected to increase from 90 Mt $H_2$/year to 530 Mt $H_2$/year, with the largest increases coming in the electricity/ heating/transportation sectors, and majority of this demand is expected to be fulfilled by these water electrolyzers [26]. However, the intrinsic variability of renewable energy sources and the need to meet hydrogen demand in a timely manner can lead to oversized GH facilities and extensive curtailment [27, 28]. On the other hand, PEM electrolyzers present a potential iridium supply chain issue, as a 1 GW-scale PEM electrolyzer would require between 5-10% of the worldwide iridium production, posing a significant bottleneck [29].

Our study falls into the broad category of an approach called "sector coupling." This approach promotes the integration of the power sector with the residential, transport, industry, and commercial/trade sectors to efficiently integrate variable energy [30]. Specifically, our current study scrutinizes the integration of solid DAC systems with GH systems through sector coupling to address existing challenges. This method ensures an optimal alignment of energy supply and demand when paired with GH production, leveraging the strengths of both technologies to enhance overall system efficiency and sustainability. Additionally, sharing power sources and energy storage devices can offer further economic benefits to the sector-coupled system.

The concept of sector coupling has been acknowledged for its potential to synergistically integrate renewable energy systems with carbon capture systems, offering economic and energy efficiency benefits [31, 32]. Rafael et al. explored the potential of sector coupling in carbon capture and storage (CCS) within industrial settings such as cement production, steel manufacture, and waste incineration [33]. By sector coupling, significant emission reductions of up to 860 Mt of $CO_2$ annually in Europe alone, equivalent to nearly 27% of European GHG emissions, are estimated. Furthermore, the integrated approach yields notable economic benefits, with cost savings of approximately 50% compared to standalone CCS technologies, emphasizing the importance of system integration in achieving net-zero and net-negative industrial sectors. Another study delved into the integration of renewable energy across Europe using the PyPSA-Eur-Sec-30 model, focusing on the implications of sector coupling and the expansion of cross-border transmission networks [34]. Their research highlights the pivotal role of technologies such as battery electric vehicles, power-to-gas units, and long-term thermal energy storage in realizing a 95% decrease in $CO_2$ emissions by 2050. Further research supports using hydrogen for long-duration energy storage in grids dependent on variable renewable energy, advocating an integrated approach for electricity and multi-technology hydrogen supply chains [35]. Favoring power-to-hydrogen pathways over power-to-gas-to-power strategies could enhance renewable energy utilization and decrease overall system costs, particularly in scenarios targeting extensive decarbonization.

Despite the growing interest in sector coupling as a strategy to enhance efficiencies of renewable energy and green chemical production, research specifically exploring the economic benefits of integrating GH with solid DAC systems remains limited. Noteworthy studies have investigated the design and operation of green methanol supply chains, which source $CO_2$ from DAC and $H_2$ from water electrolyzers, under fluctuating renewable energy supplies; however, these studies do not focus on the economic benefits of sector coupling [36, 37]. Additionally, comparative evaluations of available technology options, such as between proton exchange membrane (PEM) and alkaline electrolyzers, within the context of sector coupling are lacking. Electrolyzers may exhibit varying degrees of synergy with solid DAC in sector-coupling applications due to their differing characteristics. Our study explores the synergistic behavior of each electrolyzer technology by varying the ratio between hydrogen demand and $CO_2$ removal rates, and examining how the installation costs of these facilities influence their economic viability and synergy variations, reflecting future cost uncertainties.

Interestingly, our findings reveal that both PEM and alkaline electrolyzers can realize approximately 10% economic benefits through sector coupling with DAC systems. This advantage primarily stems from the operational flexibility of DAC systems, which, while underutilized in standalone configurations, is effectively leveraged in sector coupling. The shared use of storage devices further enhances these economic benefits. The economic advantages derived from sector coupling are significantly influenced by the specific ratio of annual hydrogen production to the system's average annual $CO_2$ reduction goal. Optimal synergy in sector coupling is achieved when the molar quantity of $CO_2$ is 1.5-3.5 times larger than that of $H_2$. This synergy indicates potential economic benefits for GH or e-fuel production, such as green methanol, with additional CCS, aiming for net-negative carbon emissions—whether the e-fuel is used as a fuel or a building resource. The economic viability of sector coupling is further enhanced when battery costs rise and DAC facility and electrolyzer costs decrease. Moreover, the substantial benefits from enhancing the operating flexibility of heat pumps are emphasized. A 10% increase in operating flexibility can reduce the total expenses of the DAC system by approximately 4%, and further boost the economic advantage of sector coupling by about 0.2% for PEM electrolyzers and 0.6% for alkaline electrolyzers.

## 2. System description

In this study, a comprehensive examination of five unique systems is undertaken: a standalone GH system, a standalone solid DAC system, and three configurations where these technologies are integrated via sector coupling. Each system harnesses wind turbines [38] and photovoltaic (PV) panels [39, 40] to convert wind and solar energy into electricity, respectively. Energy storage is facilitated through batteries [41], hydrogen tanks [42], and thermal energy storage (TES) [15], which store electricity, hydrogen, and thermal energy, correspondingly. Alkaline and PEM water electrolyzers decompose water into oxygen and hydrogen gases using electricity [24, 43], while fuel cells reconvert hydrogen into electricity [44]. Heat pumps are employed to convert electricity into thermal energy [15, 45]. Curtailment, the process of dissipating energy or idling turbines to prevent grid overload, is a critical aspect of these systems.

From a dynamic operation perspective, the minimum loads of the water electrolyzer and heat pump are of paramount importance. An increase in the minimum load of a facility can necessitate larger battery systems for electricity storage, resulting in elevated costs. The influence of these dynamics is analyzed by comparing alkaline electrolyzers and PEM electrolyzers, which have different operating characteristics. The intricacies of heat pump operations are delved into, with a particular focus on their minimum load. While operational ranges are well-established for many technologies, it is acknowledged that the minimum load for heat pumps can be influenced by a multitude of factors, rendering it less predictable.

Consequently, a sensitivity analysis is performed, investigating a range of minimum loads for heat pumps, from a conservative estimate of 40% to a more optimistic projection of 10%. This analysis allows us to scrutinize the influence of the minimum load of the heat pump on the energy system and evaluate the potential economic outcomes associated with varying degrees of operational flexibility in heat pump technology.

Given the differing current types required by each facility, efficient energy integration and operation necessitate suitable converters and bus configurations. Electrolyzers and batteries operate on direct current (DC), whereas heat pumps and DAC plants require alternating current (AC). Converter efficiencies of 98% for both DC/DC and AC/AC conversions, and 95% for AC/DC conversions are assumed in our analysis [46].

In Fig. 1 (a), the focus of the GH system is on meeting hydrogen demand promptly, with the primary energy consumption directed towards powering the water electrolyzer. Consequently, it is most efficient to configure the water electrolyzer to operate on a DC bus. Excess energy is managed through two storage strategies. Short-term energy storage is achieved by storing electricity in batteries, while $H_2$ tank serves as a longer-term energy storage solution. The system sometimes reconverts stored hydrogen into electricity via a fuel cell, ensuring that the system meets the minimum load requirement of the water electrolyzer [47]. More details on the GH system can be found in the Supplementary Materials.

In Fig. 1 (b), the solid DAC system is designed with a goal of achieving an annual average $CO_2$ capture target, making energy supply to the DAC plant its principal energy demand. As such, configuring the DAC plant to utilize an AC bus is the most efficient approach. Excess energy is stored in the form of electricity in batteries and as thermal energy in TES systems. Additionally, the system includes a heat pump, which is subject to a minimum operational requirement [48]. The supplementary materials contain further information on the DAC system.

Fig. 1 (c) depicts the integration of GH and solid DAC systems within a sector-coupled framework, featuring a hybrid DC-AC system. Purely DC or AC setups are presented in the Supplementary Materials as Fig. B. 1 and Fig. B. 2, respectively. Across all configurations, the systems are designed to promptly satisfy hydrogen demand and meet an annual $CO_2$ removal target. Each configuration employs a unique arrangement of facilities and converters to match its electrical current requirements. Specifically, the DC setup is optimized for the GH system but performs less efficiently for the DAC plant. Conversely, the AC setup offers high efficiency for the DAC plant at the expense of lower efficiency for the GH system. The hybrid DC-AC configuration achieves high efficiency within each respective bus but incurs further efficiency loss during power exchanges between DC and AC buses. This necessitates a thorough analysis to determine the most effective configuration for achieving the objectives of the sector-coupled system. In this paper, a hybrid DC-AC configured system is adopted. Further explanation is provided in Section B of the Supplementary Materials.

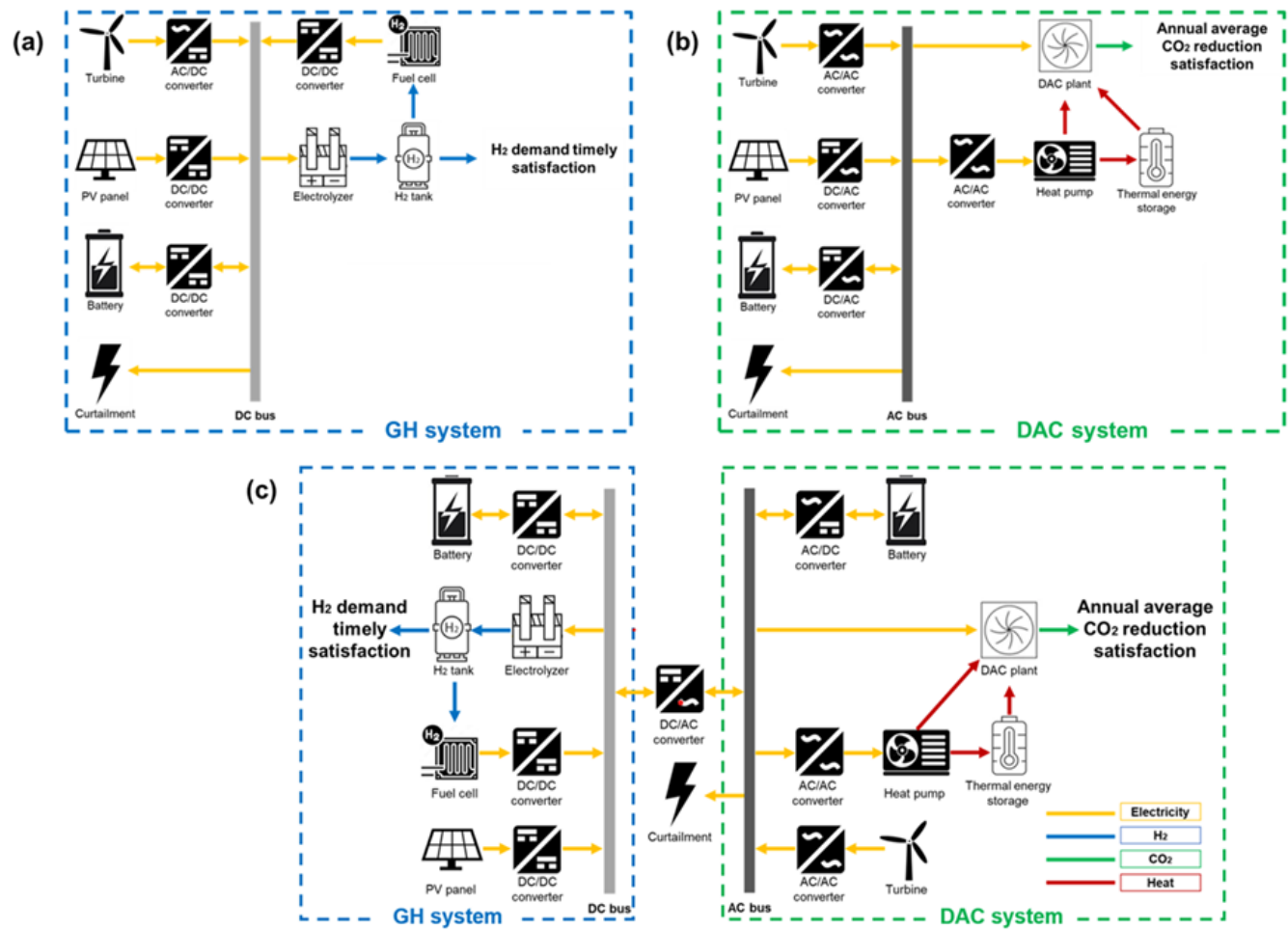

Fig. 1. (a) The GH system operating in a Direct Current (DC) configuration, designed to ensure the timely fulfillment of hydrogen demand, (b) Renewable-based Solid DAC System operating in an Alternating Current (AC) configuration, aimed at achieving an annual $CO_2$ reduction target throughout the project's duration, (c) Integrated sector coupling system of GH and solid DAC, configured to operate between DC and AC, to balance the objectives of timely satisfying hydrogen demand and achieving an annual $CO_2$ reduction target throughout the project's duration.

## 3. Methods

Our model aims to determine the optimal design that minimizes the TAC, while satisfying the annual $CO_2$ removal rates and timely $H_2$ demands, as well as adhering to the operational constraints of facilities under fluctuating renewable energy sources. The TAC encompasses expenses related to installation, operation, and maintenance as detailed in the equation below:

$$\min\ TAC = \sum_{i}\left[IC^{i} \times X_{i} \times \left(CRF^{i} + OM^{i}\right)\right] \quad \text{where} \quad CRF^{i} = \frac{r \times (1+r)^{L^{i}}}{(1+r)^{L^{i}} - 1} \quad (1)$$

In above equation, for facility $i$, $IC^{i}$ stands for its investment cost, $X_{i}$ denotes the capacity, $CRF^{i}$ signifies the capital recovery factor, and $OM^{i}$ represents the operating and maintenance cost, assumed to be proportional to the installation cost. The economic parameters of the facilities are summarized in below Table 1.

Equation (5) outlines the objective function, with constraints defined by energy balance equations and mathematical models of the facilities detailed in the Supplementary Materials, equations (A1-27). For example, the constraints for the standalone GH system are specified in equations (A1-11), whereas those for the standalone DAC system are governed by equations (A1-6) and (A12-24). The constraints for the sector-coupled system are delineated in equations (A1-24) along with the energy balance equations for each current configuration outlined in equations (A25-27).

The two-stage stochastic programming approach is employed to determine the optimal system capacity and operational strategies that effectively manage uncertainties [49, 50]. This methodology requires decision-makers to make two sequential sets of decisions. In the first stage, decisions are made about the capacity of various facilities, focusing on system design choices such as capacity and infrastructure investments without complete knowledge of future uncertainties. In the second stage, operational decisions are made in response to actual realization of uncertainties, such as variability in wind speed and solar Global Horizontal Irradiance (GHI), based on the infrastructure established in the first stage.

Further details on generating plausible weather and demand scenarios are provided in the Supplementary Materials. The IBM CPLEX solver is used to solve this linear programming problem, and the mathematical programming codes are publicly accessible via the GitHub link provided in the Data Availability section.

Table 1. Economic parameters of facilities. (projected for 2040)

| Component | Lifetime (year) | Investment cost | O&M cost (% of installation cost) | Reference |
|---|---|---|---|---|
| Wind turbine | 30 | 0.95 MM$/MW | 2.4% | [21, 51] |
| PV panel | 35 | 0.6 MM$/MW | 2% | [21, 51] |
| Battery (8 h) | 10 | 0.161 MM$/MWh | 2.5% | [21, 51] |
| PEM electrolyzer | 10 | 0.44 MM$/MW | 2% | [21, 52] |
| Alkaline electrolyzer | 10 | 0.33 MM$/MW | 2% | [21, 52] |
| Fuel cell | 5 | 0.96 MM$/MW | 2% | [44] |
| Hydrogen tank | 25 | 0.25 MM$/t $H_2$ | 1% | [53] |
| Solid DAC plant | 30 | 2.76 MM$/t $CO_2 \cdot$ h | 4% | [54] |
| Heat pump | 25 | 0.0633 MM$/MW | 0.36% | [54] |
| Thermal energy storage | 30 | 0.025 MM$/MWh$_{th}$ | 0.02% | [15] |

## 4. Evaluation metrics

The economic performance of the system is assessed using two primary metrics: the levelized cost of producing hydrogen (LCOH) and the levelized cost of DAC (LCOD), defined by equation (2) and equation (3), respectively.

$$LCOH = \frac{TAC^{GH}}{Annual\ GH\ production} \quad (2)$$

$$LCOD = \frac{TAC^{DAC}}{Annual\ CO_2\ reduction} \quad (3)$$

In the above, $TAC^{GH}$ denotes the total annualized cost (TAC) of the standalone GH system and $TAC^{DAC}$ corresponds to the TAC of the standalone solid DAC system, respectively.

For evaluating the economic synergy of sector coupling, two metrics are defined. The metric termed *Improvement* represents the difference between the TAC of the combined system and the sum of those of the standalone systems. Additionally, the metric referred to as *Synergy* denotes the ratio of the TAC difference to the sum of the standalone TACs, while ensuring timely hydrogen supply and achieving the annual CO₂ reduction target. The economic metrics are expressed as follows:

$$\text{Improvement (MM\$/yr)} = TAC^{GH} + TAC^{DAC} - TAC^{Combined} \tag{4}$$

$$\text{Synergy (\%)} = 1 - \frac{TAC^{Combined}}{TAC^{GH} + TAC^{DAC}} \tag{5}$$

In the above, $TAC^{Combined}$ indicates the TAC of the sector-coupled system. A synergy value of 0 signifies no economic advantage from the coupling, whereas a higher synergy value indicates increasing economic benefits.

## 5. Result

### 5.1. Techno-economic analysis of standalone GH and DAC system

Economic evaluation of the standalone GH and DAC systems are conducted across three locations—Antofagasta, Bundaberg, and Neom—each offering promising renewable energy potential and distinct meteorological profiles. To accommodate the seasonality and hourly fluctuation of renewable energy sources, the optimal system design is identified and economically appraised under three annual scenarios for solar and wind. Our analysis revealed that the model outcomes remain consistent beyond two annual scenarios (refer to Fig. D. 1 in the Supplementary Materials).

Fig. 2 (a) illustrates that the LCOH for PEM electrolyzer-based GH systems ranges from \$3.2/kg H₂ to \$3.6/kg H₂. The primary cost component is attributed to power sources, encompassing turbines and PV panels, followed by electrolyzers and energy storage devices. In contrast, Fig. 2 (b) shows LCOH values for alkaline electrolyzer-based GH systems range from \$3.7/kg H₂ to \$4.3/kg H₂, approximately 15% higher than PEM counterparts. Alkaline electrolyzers have lower installation costs but higher operational minimum thresholds, leading to substantial battery costs. The average annual capacity factor of the alkaline electrolyzers is approximately 55%, compared to around 40% for PEM electrolyzers, as detailed in Fig. 4 and 5. Fig. 2 (c) depicts the LCOD in various regions, ranging from \$80/t CO₂ to \$86/t CO₂, with DAC facilities as the largest cost driver, trailed by batteries and power sources. Solid DAC facilities require an annual capacity factor of approximately 90% load to achieve economic viability in standalone systems, as discussed in Figs. 4 and 5.

Moreover, sensitivity analyses are conducted by varying facility costs by ±20%. Fig. 3 (a) highlights that changes in the costs of power sources, electrolyzers, and batteries significantly impact the LCOH for PEM electrolyzer-based GH systems. Fig. 3 (b) demonstrates a different order of influence for alkaline counterparts: the costs of power sources, batteries, and electrolyzers. Fig. 3 (c) reveals that the costs of DAC, power sources, and batteries primarily affect the LCOD. Notably, in Fig. 4 (d), enhanced operating flexibility of heat pumps markedly reduces the LCOD by approximately 4% with a 10% reduction in minimum load thresholds, underscoring their role in lowering overall costs despite fixed facility costs.

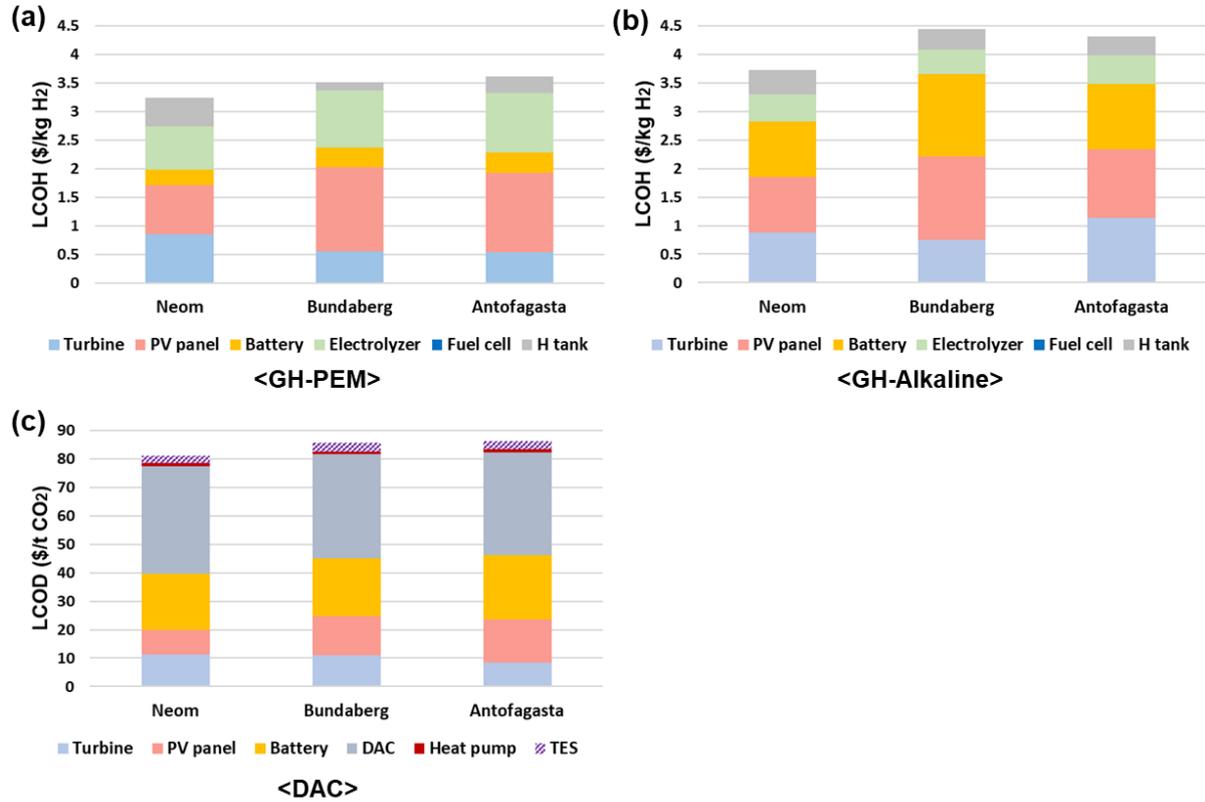

Fig. 2. Cost breakdown of standalone GH system and DAC system by varying regions: (a) PEM electrolyzer-based standalone GH system, (b) alkaline electrolyzer-based standalone GH system, (c) standalone solid DAC system. The light-blue bar corresponds to turbines, the apricot bar to PV panels, the orange bar to batteries, the green bar to electrolyzers, the blue bar to fuel cells, the grey bar to hydrogen tanks, the dark blue bar to solid DAC facilities, the red bar to heat pumps, and the purple patterned bar to the thermal energy storage devices.

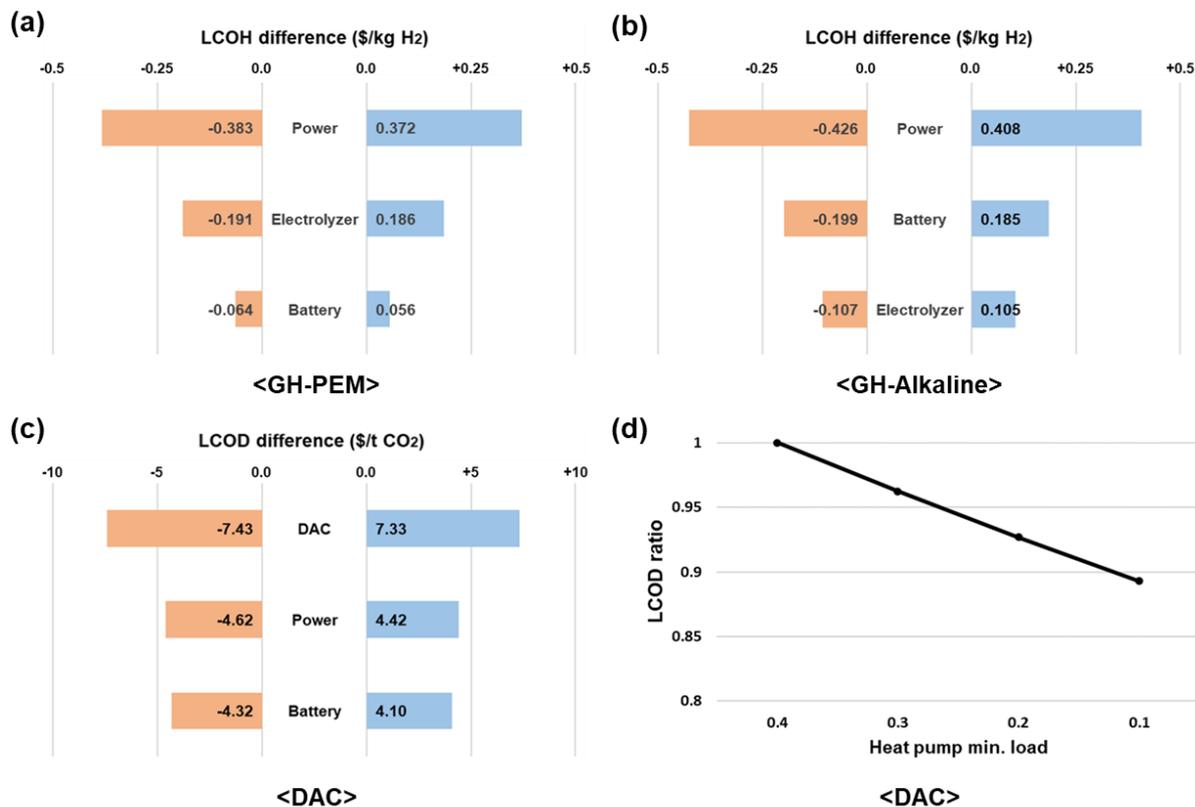

Fig. 3. Sensitivity analysis of cost variations (±20%) for different components in standalone GH and DAC systems. (a) PEM electrolyzer-based GH system, (b) alkaline electrolyzer-based GH system, (c) solid DAC system, and (d) minimum load of heat pumps in a standalone solid DAC system. The blue bars indicate +20% cost increased cases and red bars represent -20% cost reduction cases for each facility.

**5.2. Economic evaluation and synergy assessment of a sector-coupled system**

Our study evaluates the economic advantages of a sector-coupled system integrating solid DAC with GH production, compared to standalone solid DAC and GH systems. In the baseline scenario, the GH system is designed to fulfill a constant hourly hydrogen demand of 10 kt per year, matching the capacity of existing steam methane reforming plants [55]. For the solid DAC system, the economic impact of sector coupling is assessed across a range of average annual $CO_2$ removal rate, from 0.1 Mt to 1.3 Mt, in 0.2 Mt increments. In molar terms, 0.2 Mt of $CO_2$ is approximately equivalent to 10 kt of $H_2$.

TAC breakdown is explored for both standalone and sector-coupled systems, highlighting the economic enhancements achieved through sector coupling by approximately 5% at a $CO_2$ removal rate of 1 Mt/yr, as detailed in Tables 2 and 3. Notably, the fuel cell, which performs the power-to-gas-to-power process, is not employed in either the standalone or sector-coupled systems.

From a $CO_2$ removal perspective, the sector-coupled system increases DAC facility installations while reducing capital investments in power sources, batteries, and TES. A standalone DAC system requires substantial power sources and batteries to reduce investment in DAC facilities. In contrast, a sector-coupled system can operate with wider operational variations depending on energy availability, reducing the need for additional power sources and batteries. Furthermore, the optimal capacity of TES decreases as the larger DAC facility, with the same annual $CO_2$ removal rate, can handle excess thermal energy more efficiently.

From a hydrogen production perspective, the size of electrolyzer increases while the capacity of $H_2$ tank is reduced. The average monthly capacify factor of PEM electrolyzers is 10-20% lower than that of alkaline electrolyzers. The higher minimum load of alkaline electrolyzers limits the feasibility of large installations, whereas PEM electrolyzers face fewer constraints, allowing for larger installations. Consequently, in a sector-coupled system, it is easier to meet hydrogen demand, leading to smaller $H_2$ tank installations. The operational pattern is detailed in the hourly operational decision graph in the Supplementary Material (see Figs. F.2-4 in the Supplementary Material).

Initially, it was anticipated that alkaline electrolyzer-based sector-coupled systems would show greater improvement and synergy in battery TAC components than PEM counterparts, due to the substantial battery installations required for standalone DAC systems and alkaline electrolyzer-based GH systems. However, our observations indicate that PEM electrolyzers achieve more significant benefits: a $3.23 million/yr improvement and 23.9% synergy in batteries through sector coupling, compared to $2.74 million/yr improvement and 13.6% synergy for alkaline counterparts. The higher operating flexibility of PEM electrolyzers enables them to effectively utilize the DAC system's flexibility, particularly in response to the short-term variability of renewable energy sources, which facilitates efficient hourly operation (as detailed in Figs. F.2-4 in the Supplementary Material). Conversely, the sector coupling strategy does not significantly reduce the need for substantial energy storage to meet the minimum load requirements of alkaline electrolyzers. This highlights that dynamically leveraging the operational flexibility of DAC and PEM electrolyzers provides a considerable advantage over merely sharing facilities.

Fig 4. Illustrates the TAC breakdown of sector-coupled systems across different regions and $CO_2$ removal rates. As in our previous observation that PEM electrolyzers demonstrated approximately 15% greater cost-effectiveness compared to their alkaline counterparts. Additionally, a PEM electrolyzer-based sector-coupled system exhibits higher economic synergy than its alkaline counterpart. As a result, alkaline electrolyzer-based sector-coupled systems incur higher costs due to the need for larger installations of power sources, batteries, and fuel cells.

Fig. 5 (a) underscores the improvement between $2 million and $9 million per year when PEM electrolyzers and DAC are sector-coupled, increasing steadily as the $CO_2$ removal rate rises. The highest synergy is observed when the $CO_2$ removal rate is between 0.3-0.7 Mt/yr. Similarly, Fig. 5 (b) illustrates that the improvement within the sector-coupled scenario of alkaline electrolyzers and DAC ranges between $2 million and $9 million per year, following a similar trend of increasing improvement as $CO_2$ removal rate rises. The peak synergy occurs at $CO_2$ removal rates of 0.3-0.7 Mt/yr. These findings suggest that incorporating a sector-coupled system enhances economic benefits, potentially offering significant advantages for GH or e-fuel production, such as green methanol, with additional CCS to achieve net negative carbon emissions, whether the e-fuel is utilized as an energy source or as a building material.

Table 2. TAC breakdown (MM$/yr) of PEM electrolyzer-based systems. This table analyzes the averaged improvement (MM$/yr) and synergy (%) achieved through sector coupling at a carbon removal rate of 1 Mt/yr across all regions. It compares the TAC of a standalone GH system, a standalone DAC system, the combined total of both standalone systems (Combined), and a sector-coupled system.

|  | Power Sources | Battery | Electrolyzer | Fuel Cell | $H_2$ tank | DAC | Heat Pump | TES | Total |
|---|---|---|---|---|---|---|---|---|---|
| Standalone GH TAC | 18.85 | 3.26 | 9.31 | 0.01 | 3.12 | 0.00 | 0.00 | 0.00 | 34.55 |
| Standalone DAC TAC | 22.66 | 20.96 | 0.00 | 0.00 | 0.00 | 36.83 | 0.98 | 2.86 | 84.29 |
| Combined TAC | 41.51 | 24.22 | 9.31 | 0.01 | 3.12 | 36.83 | 0.98 | 2.86 | 118.84 |
| Sector-coupled TAC | 39.08 | 21.80 | 8.64 | 0.00 | 1.45 | 38.01 | 0.96 | 2.27 | 112.22 |
| Improvement (MM$/yr) | 2.43 | 2.42 | 0.67 | 0.00 | 1.67 | -1.18 | 0.02 | 0.59 | 6.62 |
| Synergy (%) | 5.72 | 9.90 | 6.10 | - | 47.48 | -3.19 | 1.97 | 22.37 | 5.56 |

Table 3. TAC breakdown (MM$/yr) of alkaline electrolyzer-based systems. This table analyzes the averaged improvement (MM$/yr) and synergy (%) achieved through sector coupling at a carbon removal rate of 1 Mt/yr across all regions. It compares the TAC of a standalone GH system, a standalone DAC system, the combined total of both standalone systems (Combined), and a sector-coupled system.

|  | Power Sources | Battery | Electrolyzer | Fuel Cell | $H_2$ tank | DAC | Heat Pump | TES | Total |
|---|---|---|---|---|---|---|---|---|---|
| Standalone GH TAC | 21.35 | 11.86 | 4.65 | 0.00 | 3.73 | 0.00 | 0.00 | 0.00 | 41.59 |
| Standalone DAC TAC | 22.66 | 20.96 | 0.00 | 0.00 | 0.00 | 36.83 | 0.98 | 2.86 | 84.29 |
| Combined TAC | 44.01 | 32.83 | 4.65 | 0.00 | 3.73 | 36.83 | 0.98 | 2.86 | 125.88 |
| Sector-coupled TAC | 40.40 | 31.42 | 4.81 | 0.00 | 1.26 | 37.88 | 0.89 | 2.77 | 119.43 |
| Improvement (MM$/yr) | 3.60 | 1.41 | -0.15 | 0.00 | 2.47 | -1.05 | 0.08 | 0.09 | 6.45 |
| Synergy (%) | 7.94 | 4.42 | -3.09 | - | 66.92 | -2.86 | 8.47 | 3.17 | 5.11 |

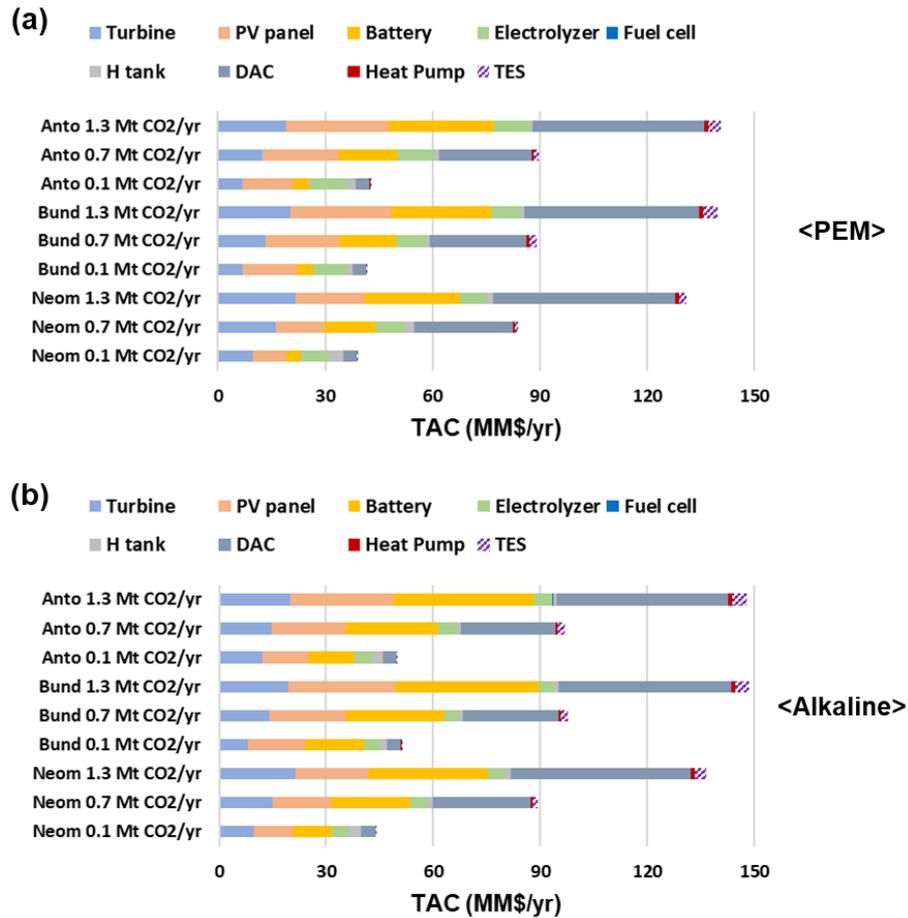

Fig. 4. TAC breakdown of sector-coupled systems across different regions and $CO_2$ removal rates: (a) PEM electrolyzer-based sector-coupled system, (b) alkaline electrolyzer-based sector-coupled system. The light-blue bar corresponds to turbines, the red bar represents PV panels, the orange bar denotes batteries, the green bar indicates electrolyzers, the blue bar indicates fuel cells, the grey bar denotes hydrogen tanks, the dark blue bar represents solid DAC facilities, the dark red bar indicates heat pumps, and the purple patterned bar denotes the thermal energy storage devices.

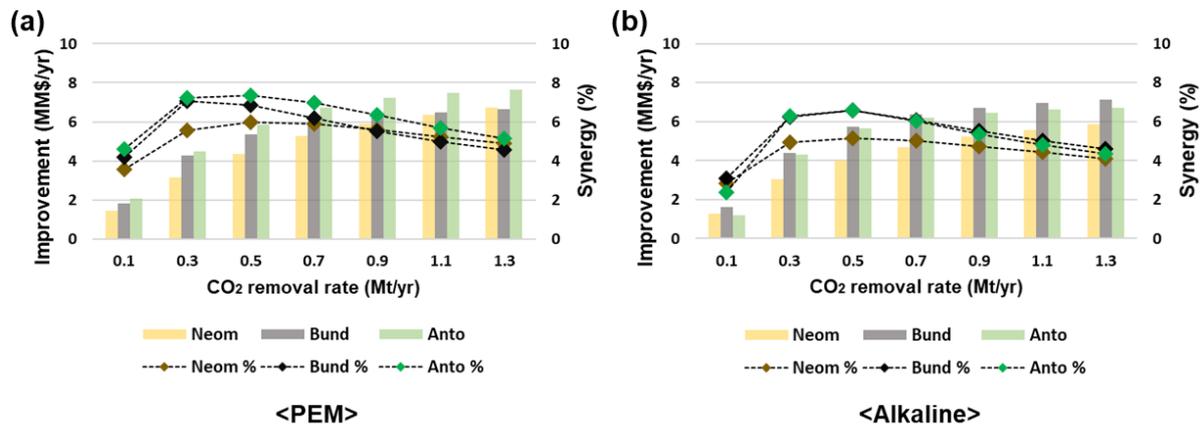

Fig. 5. Improvement (MM$/yr) and synergy (%) of sector-coupled system across different regions by varying $CO_2$ removal rate: (a) PEM electrolyzer-based sector-coupled system, (b) alkaline electrolyzer-based sector-coupled system. The yellow bar indicates improvement at Neom, the grey bar denotes improvement at Bundaberg, and the green bar represents improvement at Antofagasta. The red dashed line denotes synergy at Neom, the black dashed line represents synergy at Bundaberg, and the green dashed line indicates synergy at Antofagasta.

### 5.3. Techno-economic analysis of sector-coupled system and their synergy variation

The examination of how variations in facility costs impact the economic feasibility and synergies of sector-coupled systems is conducted, given uncertainties in future renewable technology cost projections. This analysis aims to highlight potential synergies within sector-coupled systems across diverse future scenarios and scrutinize the economic implications and synergistic effects of our proposed sector coupling model amid substantial fluctuations in key facility costs. Specifically, the economic variations are assessed when costs of individual facilities—such as batteries, DAC, electrolyzers, and power source, which are major cost drivers—increase by 20% from the baseline scenario. Their impacts are evaluated by analyzing the TAC difference (MM$/yr) and synergy difference (%).

In both sector-coupled systems, a noticeable trend emerges where the TAC difference escalates with increasing $CO_2$ removal rates from 0.1 Mt/yr to 1.3 Mt/yr, as depicted in Figs. 6 (a) and (b). A 20% increase in power source costs results in a TAC increment from approximately $4.1 million/yr at a $CO_2$ removal rate of 0.1 Mt/yr to about $8.8 million/yr at 1.3 Mt/yr for PEM electrolyzer-based systems. In contrast, alkaline counterparts show an increase from $4.4 million/yr to about $9.2 million/yr. A 20% rise in DAC facility costs leads to a TAC increase from around $0.8 million/yr to approximately $10 million/yr in both systems. Additionally, a 20% hike in battery costs results in a TAC for PEM electrolyzer-based systems rising from roughly $0.8 million/yr to about $4.3 million/yr, while for alkaline counterparts, it increases from $2 million/yr to approximately $6 million/yr. Notably, a 20% rise in electrolyzer costs consistently results in a TAC increase of about $1.9 million/yr for PEM electrolyzers and

around $1.1 million/yr for alkaline electrolyzers, irrespective of the $CO_2$ removal rate. The PEM electrolyzer-based system exhibits approximately twice the sensitivity to electrolyzer and battery costs compared to alkaline counterparts, while showing similar variations for DAC facility and power sources.

When assessing the synergy difference, a clear contrast emerges between the systems. Fig. 6 (c) shows that, for PEM electrolyzer cases, as $CO_2$ removal rates increase from 0.1 Mt/yr to 1.3 Mt/yr, the synergy difference due to the 20% higher battery costs escalates from +0.34% to +0.8%. In contrast, Fig. 6 (d) indicates a more modest rise from +0.17% to +0.53% for alkaline electrolyzer cases under the same condition. This demonstrates that PEM electrolyzers benefit more significantly from a 20% hike in battery costs, indicating the greater advantage of utilizing the operational flexibility of DAC and PEM electrolyzer compared to their alkaline counterparts. This result aligns with the findings in Tables 1 and 2, showing that the improvement in the battery cost component is greater in PEM electrolyzer-based sector-coupled systems compared to their alkaline counterparts.

Regarding a 20% increase in DAC facility costs, both systems exhibit a similar trajectory, with a synergy difference of approximately -0.16% at a $CO_2$ removal rate of 0.1 Mt/yr and about -0.6% at rates between 0.7 and 1.3 Mt/yr for PEM electrolyzer cases, and -0.12% at 0.1 Mt/yr rate to about -0.55% at rates between 0.7 and 1.3 Mt/yr for alkaline electrolyzer cases. Typically, the high installation costs of DAC units necessitate opting for smaller capacities operating at higher average rates to maintain cost-effectiveness. Consequently, as DAC unit installation costs decrease in the future, the benefits of sector coupling will become more evident. Reduced installation costs encourage the expansion of DAC's optimal capacity, facilitating more dynamic operation of the DAC system.

Similar to the DAC facility case, the advantages of sector coupling diminish as electrolyzer costs rise, leading to smaller installations and higher average operating rates of electrolyzers. However, there is a significant disparity in how the two systems respond to rising electrolyzer costs: in the sector-coupled system employing a PEM electrolyzer, the synergy decline stands at -0.2% at a $CO_2$ removal rate of 0.1 Mt/yr and -0.1% at 1.3 Mt/yr. Conversely, the alkaline electrolyzer counterpart shows the largest synergy decline of -0.18% at 0.5 Mt/yr. This difference reflects the different characteristics of each electrolyzer, leading to varying trend in synergy difference with increasing $CO_2$ removal rates.

An intriguing finding emerges when the cost of the power source facility (turbine, PV panel) is increased by 20%. In both systems, a negative synergy difference is observed when the $CO_2$ removal rate is below 0.5 Mt/yr, but a positive synergy difference emerges once the threshold is surpassed. Sector coupling derives economic advantages from two factors: first, it allows DAC to operate at a broader monthly capacity factor; second, it prevents overinvestment by sharing power sources and batteries. When the power source becomes more expensive, the first advantage of sector coupling diminishes, while the second advantage increases. At lower $CO_2$ removal rates, the reduction in DAC's capacity factor variability (the first reason) has a more pronounced effect, resulting in a negative synergy difference. As the $CO_2$ removal rate rises, the impact of shared facilities (the second reason) becomes more dominant, leading to a positive synergy difference. Consequently, a negative synergy difference is observed below a $CO_2$ removal rate of 0.5 Mt/yr with the synergy difference turning positive beyond this threshold.

Interestingly, at each $CO_2$ removal rate, the sum of synergy differences for a 20% increase in the installed price of a major facility nearly equals zero. This indicates that the synergistic effects maintain notable consistency compared to those observed in the baseline scenario, regardless of overarching technological advancements or regressions. Essentially, the benefits of sector coupling persist despite the evolution of renewable energy technology.

On the other hand, a sensitivity analysis is conducted by varying the minimum loads for heat pumps from 40% to 10%. This analysis allowed us to explore how the operating dynamics of the heat pump influence the energy system and assess the associated economic outcomes. Figs. 7 (a) and (b) show that as the minimum load of the heat pump decreases leading to enhanced operational flexibility, the TAC also decreases. The TAC ratio, which compares the TAC when the minimum load of the heat pump is reduced to that of the base case (40% minimum load), illustrates this trend. A lower TAC ratio signifies a greater reduction in TAC compared to the base case. In both systems, as the $CO_2$ removal rate increases, the reduction in TAC ratio becomes more significant with a decreasing minimum load of the heat pump. For example, at a $CO_2$ removal rate of 0.1 Mt/yr, a 10% increase in heat pump operating flexibility results in approximately a 1% decrease in the TAC ratio. This reduction increases to about 2.7% for PEM electrolyzer case and approximately 3.0% for alkaline counterparts at a $CO_2$ removal rate of 1.3 Mt/yr.

Examining Figs. 7 (c) and (d), the overall synergy trend in the base case remains consistent, albeit with slight variations. Specifically, in the sector-coupled system employing a PEM electrolyzer, synergy increases by up to 1% when the $CO_2$ removal rate is below 0.9 Mt/yr due to the improved operating flexibility of the heat pump. Conversely, for $CO_2$ removal rates exceeding 1.1 Mt/yr, synergy experiences a minor decline as the operating flexibility of the heat pump increases. In contrast, in the system with an alkaline electrolyzer, while the overall trend aligns similarly, the synergy attributed to the heat pump's operating flexibility gradually diminishes as the $CO_2$ removal rate rises, with the variation in synergy is about twice as large as that seen in the PEM electrolyzer counterparts. This observation underscores the complex interplay within the components of the sector-coupled system and emphasizes the critical need for system-wide optimization to fully harness the collective advantages.

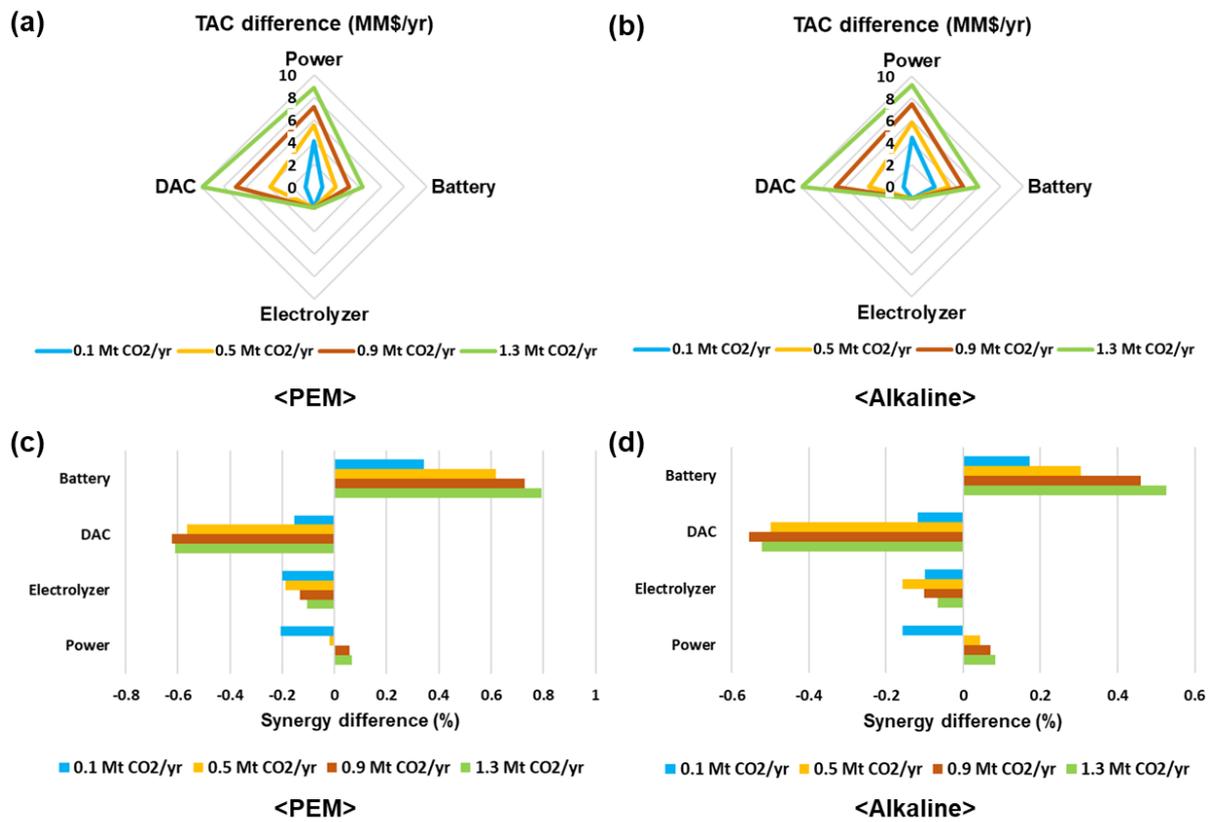

Fig. 6. Variation in TAC difference (MM$/yr) and synergy difference (%) resulting from a 20% increased cost in the cost of each facility, while varying $CO_2$ removal rate in the sector-coupled system: (a) TAC difference in PEM electrolyzer-based sector-coupled system, (b) TAC difference in alkaline electrolyzer-based sector-coupled system, (c) Synergy difference in PEM electrolyzer-based sector-coupled system, (d) Synergy difference in alkaline electrolyzer-based sector-coupled system. Color coding for $CO_2$ removal rate is as follows: light-blue represents 0.1 Mt $CO_2$/yr, yellow for 0.5 Mt $CO_2$/yr, brown for 0.9 Mt $CO_2$/yr, and light-green for 1.3 Mt $CO_2$/yr.

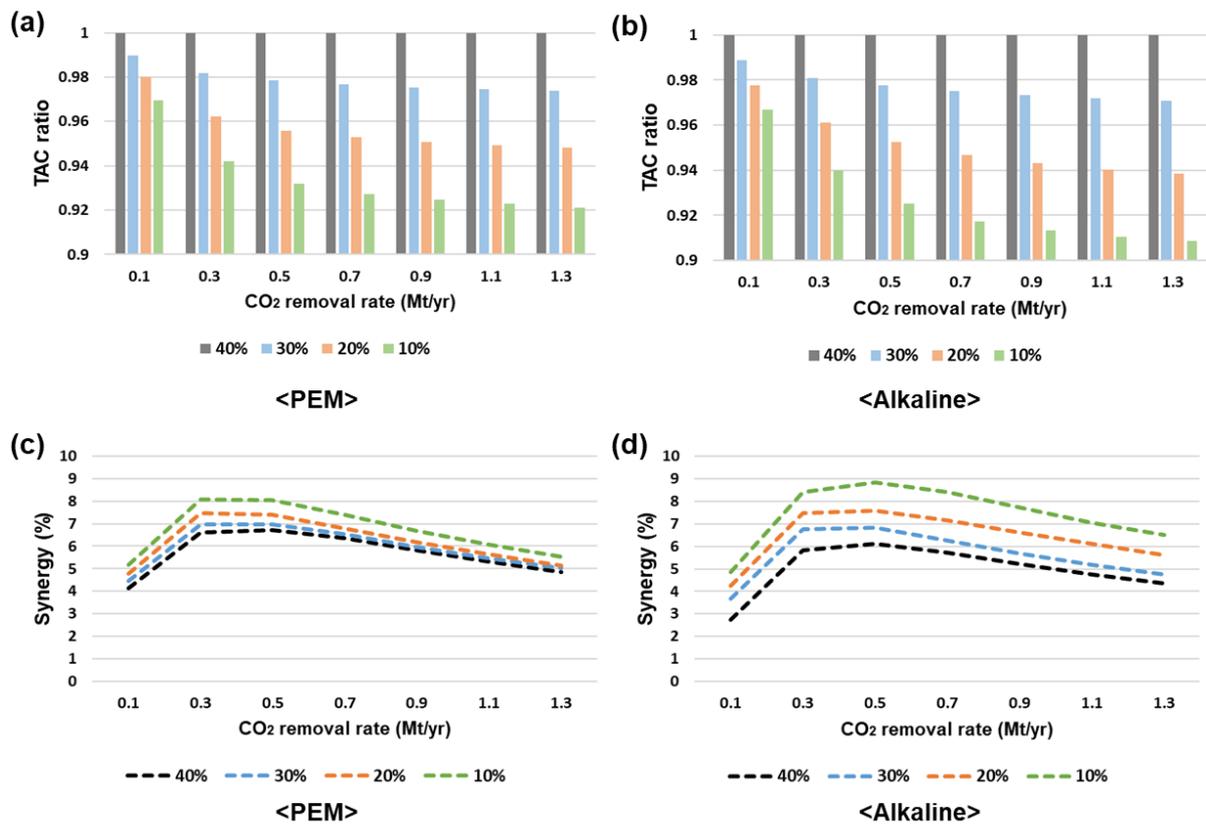

Fig. 7. TAC ratio referenced in the base case across different heat pump minimum loads by varying $CO_2$ removal rate for (a) PEM electrolyzer-based sector-coupled system and (b) alkaline electrolyzer-based sector-coupled system. Synergy (%) across different heat pump minimum loads by varying $CO_2$ removal rate for (c) PEM electrolyzer-based sector-coupled system and (d) alkaline electrolyzer-based sector-coupled system. Color coding for the heat pump minimum loads is as follows: black indicates the base case, blue represents a 30% minimum load, red denotes a 20% minimum load, and green represents a 10% minimum load.

## 6. Conclusion

Our study advocates for the sector coupling strategy integrating GH production with solid DAC systems, highlighting its significant economic benefits. This research presents an optimization-based approach to address the dual challenges of reducing anthropogenic $CO_2$ emissions and meeting timely $H_2$ demand. The synergistic behavior of different electrolyzer technologies is also explored by varying the ratio between hydrogen demand and $CO_2$ removal rates, with an analysis of how the installation costs of these facilities impact their economic feasibility and synergy, reflecting future cost uncertainties. Furthermore, the importance of operating flexibility in heat pump for economic feasibility are emphasized.

The key findings from our analysis include:

- **Economic Benefits:** Integrating GH production with solid DAC systems yields substantial economic benefits, demonstrating approximately a 10% improvement across the board.
- **Investment Reduction:** The benefits of sector coupling are primarily driven by reduced investments in power sources, batteries, $H_2$ tanks, and TES, while simultaneously increasing the variation in DAC facilities' monthly capacity factor.
- **Operating Dynamics:** A different operational dynamic emerges when hydrogen is the main demand source. This contrasts with scenarios where electricity is the primary demand source and hydrogen is typically used to bridge long-term energy gaps.
- **Influence of Production to Reduction Ratio:** The economic viability of sector coupling is significantly influenced by the ratio of GH production to the system's annual $CO_2$ reduction goal.
- **Optimal Synergy:** Optimal synergy in sector coupling is achieved with a $CO_2$-to-$H_2$ molar ratio of 1.5-3.5 in both systems, offering economic advantages for GH or e-fuel production with additional carbon capture.
- **Sensitivities of Cost Benefits:** Cost benefits of sector coupling are enhanced by increased battery costs and decreased costs for DAC facilities and electrolyzers, with the impact of power source costs varying based on $CO_2$ removal rates.
- **Consistency of Benefits:** Sector coupling consistently enhances efficiency and sustainability, offering enduring benefits regardless of future developments in renewable energy technology.
- **Benefits of Enhanced Operational Flexibility of Heat Pumps:** A 10% increase in the operational flexibility of heat pumps improves the economic feasibility of standalone DAC systems by 4% and further enhances the economic synergy driven by sector coupling.

In summary, our study proposes the sector coupling strategy of integrating GH production with solid DAC systems and provides insights into its economic advantages, contributing to the discourse on sustainable energy solutions. These findings are essential for policymakers, industry stakeholders, and researchers dedicated to advancing a low-carbon future. The sector coupling approach particularly favors alkaline electrolyzers over PEM electrolyzers, underscoring its potential in shaping future energy landscapes.

**Conflicts of interest**

There are no conflicts of interest to declare

**Acknowledgements**

The authors acknowledge that this work was supported by the KAIST-Aramco $CO_2$ Management Center.

**References**

1. Liu, S., et al., *Direct air capture of CO2 via cyclic viologen electrocatalysis.* Energy & Environmental Science, 2024. **17**(3): p. 1266-1278.
2. Sharifian, R., et al., *Electrochemical carbon dioxide capture to close the carbon cycle.* Energy & Environmental Science, 2021. **14**(2): p. 781-814.
3. Holmes, H.E., M.J. Realff, and R.P. Lively, *Water management and heat integration in direct air capture systems.* Nature Chemical Engineering, 2024. **1**(3): p. 208-215.
4. Jacobson, M.Z., *The health and climate impacts of carbon capture and direct air capture.* Energy & Environmental Science, 2019. **12**(12): p. 3567-3574.
5. Küng, L., et al., *A roadmap for achieving scalable, safe, and low-cost direct air carbon capture and storage.* Energy & Environmental Science, 2023. **16**(10): p. 4280-4304.
6. Breyer, C., et al., *Direct Air Capture of CO2: A Key Technology for Ambitious Climate Change Mitigation.* Joule, 2019. **3**(9): p. 2053-2057.
7. Bui, M., et al., *Carbon capture and storage (CCS): the way forward.* Energy & Environmental Science, 2018. **11**(5): p. 1062-1176.
8. Bachman, E., et al., *Rail-based direct air carbon capture.* Joule, 2022. **6**(7): p. 1368-1381.
9. Young, J., et al., *The impact of binary water–CO2 isotherm models on the optimal performance of sorbent-based direct air capture processes.* Energy & Environmental Science, 2021. **14**(10): p. 5377-5394.
10. Sabatino, F., et al., *A comparative energy and costs assessment and optimization for direct air capture technologies.* Joule, 2021. **5**(8): p. 2047-2076.
11. Holmes, H.E., et al., *Tuning sorbent properties to reduce the cost of direct air capture.* Energy & Environmental Science, 2024.
12. Breyer, C., M. Fasihi, and A. Aghahosseini, *Carbon dioxide direct air capture for effective climate change mitigation based on renewable electricity: a new type of energy system sector coupling.* Mitigation and Adaptation Strategies for Global Change, 2019. **25**(1): p. 43-65.
13. Young, J., et al., *Process-informed adsorbent design guidelines for direct air capture.* Chemical Engineering Journal, 2023. **456**.
14. Erans, M., et al., *Direct air capture: process technology, techno-economic and socio-political challenges.* Energy & Environmental Science, 2022. **15**(4): p. 1360-1405.
15. Yang, L. and X. Wu, *Net-zero carbon configuration approach for direct air carbon capture based integrated energy system considering dynamic characteristics of CO2 adsorption and desorption.* Applied Energy, 2024. **358**.
16. Shirizadeh, B., et al., *Towards a resilient and cost-competitive clean hydrogen economy: the future is green.* Energy & Environmental Science, 2023. **16**(12): p. 6094-6109.
17. Liu, H., et al., *Pathway toward cost-effective green hydrogen production by solid oxide electrolyzer.* Energy & Environmental Science, 2023. **16**(5): p. 2090-2111.
18. Staffell, I., et al., *The role of hydrogen and fuel cells in the global energy system.* Energy & Environmental Science, 2019. **12**(2): p. 463-491.

19. Park, J., et al., *Techno-economic analysis of solar powered green hydrogen system based on multi-objective optimization of economics and productivity.* Energy Conversion and Management, 2024. **299**.
20. Won, W., et al., *Design and operation of renewable energy sources based hydrogen supply system: Technology integration and optimization.* Renewable Energy, 2017. **103**: p. 226-238.
21. Terlouw, T., et al., *Large-scale hydrogen production via water electrolysis: a techno-economic and environmental assessment.* Energy & Environmental Science, 2022. **15**(9): p. 3583-3602.
22. Voronova, A., et al., *Systematic degradation analysis in renewable energy-powered proton exchange membrane water electrolysis.* Energy & Environmental Science, 2023. **16**(11): p. 5170-5184.
23. Hu, K., et al., *Comparative study of alkaline water electrolysis, proton exchange membrane water electrolysis and solid oxide electrolysis through multiphysics modeling.* Applied Energy, 2022. **312**.
24. Buttler, A. and H. Spliethoff, *Current status of water electrolysis for energy storage, grid balancing and sector coupling via power-to-gas and power-to-liquids: A review.* Renewable and Sustainable Energy Reviews, 2018. **82**: p. 2440-2454.
25. Feng, Q., et al., *A review of proton exchange membrane water electrolysis on degradation mechanisms and mitigation strategies.* Journal of Power Sources, 2017. **366**: p. 33-55.
26. Sun, E., et al., *Requirements for CO2-free hydrogen production at scale.* Joule, 2024. **8**(6): p. 1539-1543.
27. Park, J., et al., *Green hydrogen to tackle the power curtailment: Meteorological data-based capacity factor and techno-economic analysis.* Applied Energy, 2023. **340**.
28. Joos, M. and I. Staffell, *Short-term integration costs of variable renewable energy: Wind curtailment and balancing in Britain and Germany.* Renewable and Sustainable Energy Reviews, 2018. **86**: p. 45-65.
29. Dubouis, N., et al., *Alkaline electrolyzers: Powering industries and overcoming fundamental challenges.* Joule, 2024. **8**(4): p. 883-898.
30. Ramsebner, J., et al., *The sector coupling concept: A critical review.* WIREs Energy and Environment, 2021. **10**(4).
31. Bernath, C., G. Deac, and F. Sensfuß, *Impact of sector coupling on the market value of renewable energies – A model-based scenario analysis.* Applied Energy, 2021. **281**.
32. Gea-Bermúdez, J., et al., *The role of sector coupling in the green transition: A least-cost energy system development in Northern-central Europe towards 2050.* Applied Energy, 2021. **289**.
33. Castro-Amoedo, R., et al., *On the role of system integration of carbon capture and mineralization in achieving net-negative emissions in industrial sectors.* Energy & Environmental Science, 2023. **16**(10): p. 4356-4372.
34. Brown, T., et al., *Synergies of sector coupling and transmission reinforcement in a cost-optimised, highly renewable European energy system.* Energy, 2018. **160**: p. 720-739.
35. He, G., et al., *Sector coupling via hydrogen to lower the cost of energy system decarbonization.* Energy & Environmental Science, 2021. **14**(9): p. 4635-4646.
36. Fasihi, M. and C. Breyer, *Global production potential of green methanol based on variable renewable electricity.* Energy & Environmental Science, 2024. **17**(10): p. 3503-3522.

37. Fulham, G.J., P.V. Mendoza-Moreno, and E.J. Marek, *Managing intermittency of renewable power in sustainable production of methanol, coupled with direct air capture.* Energy & Environmental Science, 2024.
38. Chauhan, A. and R.P. Saini, *A review on Integrated Renewable Energy System based power generation for stand-alone applications: Configurations, storage options, sizing methodologies and control.* Renewable and Sustainable Energy Reviews, 2014. **38**: p. 99-120.
39. Bukar, A.L., et al., *A rule-based energy management scheme for long-term optimal capacity planning of grid-independent microgrid optimized by multi-objective grasshopper optimization algorithm.* Energy Convers Manag, 2020. **221**: p. 113161.
40. Bukar, A.L., C.W. Tan, and K.Y. Lau, *Optimal sizing of an autonomous photovoltaic/wind/battery/diesel generator microgrid using grasshopper optimization algorithm.* Solar Energy, 2019. **188**: p. 685-696.
41. Liu, T., et al., *Techno-economic feasibility of solar power plants considering PV/CSP with electrical/thermal energy storage system.* Energy Conversion and Management, 2022. **255**.
42. Andersson, J. and S. Grönkvist, *Large-scale storage of hydrogen.* International Journal of Hydrogen Energy, 2019. **44**(23): p. 11901-11919.
43. Grigoriev, S.A., et al., *Current status, research trends, and challenges in water electrolysis science and technology.* International Journal of Hydrogen Energy, 2020. **45**(49): p. 26036-26058.
44. Nayak-Luke, R.M. and R. Bañares-Alcántara, *Techno-economic viability of islanded green ammonia as a carbon-free energy vector and as a substitute for conventional production.* Energy & Environmental Science, 2020. **13**(9): p. 2957-2966.
45. Wiegner, J.F., et al., *Optimal Design and Operation of Solid Sorbent Direct Air Capture Processes at Varying Ambient Conditions.* Industrial & Engineering Chemistry Research, 2022. **61**(34): p. 12649-12667.
46. Kim, S., et al., *Multi-period, multi-timescale stochastic optimization model for simultaneous capacity investment and energy management decisions for hybrid Micro-Grids with green hydrogen production under uncertainty.* Renewable and Sustainable Energy Reviews, 2024. **190**.
47. Niblett, D., et al., *Review of next generation hydrogen production from offshore wind using water electrolysis.* Journal of Power Sources, 2024. **592**.
48. Meesenburg, W., et al., *Optimizing control of two-stage ammonia heat pump for fast regulation of power uptake.* Applied Energy, 2020. **271**.
49. You, F., J.M. Wassick, and I.E. Grossmann, *Risk management for a global supply chain planning under uncertainty: Models and algorithms.* AIChE Journal, 2009. **55**(4): p. 931-946.
50. Yeh, K., et al., *Two stage stochastic bilevel programming model of a pre-established timberlands supply chain with biorefinery investment interests.* Computers & Chemical Engineering, 2015. **73**: p. 141-153.
51. Vimmerstedt, L., et al., *Annual Technology Baseline: The 2022 Electricity Update.* 2022, National Renewable Energy Lab.(NREL), Golden, CO (United States).
52. Böhm, H., et al., *Projecting cost development for future large-scale power-to-gas implementations by scaling effects.* Applied Energy, 2020. **264**.
53. Ma, N., et al., *Large scale of green hydrogen storage: Opportunities and challenges.* International Journal of Hydrogen Energy, 2024. **50**: p. 379-396.

54. Fasihi, M., O. Efimova, and C. Breyer, *Techno-economic assessment of CO2 direct air capture plants.* Journal of Cleaner Production, 2019. **224**: p. 957-980.
55. Reed, J., Emily Dailey, Brendan Shaffer, Blake Lane, Robert Flores, Amber Fong, G. Scott Samuelsen., *Roadmap for the Deployment and Buildout of Renewable Hydrogen Production Plants in California*, C.E. Commission, Editor. 2020.